# Négociation de spectre dans les réseaux de radio cognitive

**Rapport de recherche - Juin 2013**

Laboratoire de Télécommunications de Tlemcen (LTT)

Université Abou Bekr Belkaid Tlemcen

Ibtissam Larbi et Badr Benmammar

**Table des matières**

**Listes des figures**

**Listes des tableaux**

**Liste des abréviations**

| Acronyme | Signification |
|---|---|
| ACC | Agent Communication Chanel |
| ACL | Agent communication Language |
| AMD | Agent Management System |
| AP | Access Point |
| API | Application Programme Interface |
| DF | Director Facilitator |
| FIPA | Fountion for Intelligent Physical Agents |
| IA | Intelligence Artificielle |
| IAD | Intelligence Artificielle Distrubuée |
| JADE | Java Agent Development Framework |
| KTH | Kungliga Tekniska hogskolan (Institut royal de technologie) |
| LGPL | Lesser General Public Licence |
| MAC | Medium Access Control |
| PU | Primary User |
| RC | Radio Cognitive |
| RF | Radio Freqency |
| RL | Radio Logicielle |
| SDR | Radio Logicielle Restreine |
| SMA | Système Multi Agents |
| SP | Sensory Perception |
| SU | Secondary User |
| WRAN | Wireless Regional Access Networks |

**Introduction générale**

La demande croissante de la communication sans-fil introduit le défi de l'utilisation efficace du spectre. Pour relever ce défi L'IEEE 802.22 est un nouveau groupe de travail du comité de normalisation WRAN. La radio cognitive est apparue comme une technologie clé, qui perme un accès opportuniste au spectre pour répondre directement au besoin applicatifs des utilisateurs de la radio cognitive.

Dans le cas de l'utilisation des bandes sans licence, nous savons qu'il y a des interactions entre les entités c'est-à-dire entre les PUs (utilisateurs primaires) et le les SUs (utilisateurs secondaires), le PU est celui qui a une licence sur le spectre, le SU et l'utilisateur opportuniste. Un problème de l'encombrement causé par le manque de ressources disponibles pour les SUs est apparu. Pour résoudre ce problème, les réseaux de radio cognitive utilisent l'accès dynamique au spectre. Dans ce RAPPORT, nous proposons une technique basée sur la négociation dans le cadre d'un système multi agents (SMA) qui est particulièrement adaptés pour proposer des solutions réactives à des problèmes complexes comme la négociation du spectre. Nous avons implémente la solution proposée avec JADE (Java Agent Development Framework), nous avons également fait l'évaluation dans la solution proposé pour montrer son intérêt.

Ce rapport est organisé comme suit :
Dans le premier chapitre introduit plusieurs concept comme la radio logicielle, la radio logicielle restreinte et la radio cognitive. Nous allons présenter les caractéristiques principales de la radio cognitive et leur fonctionnement, le deuxième chapitre se concentre sur les diverses méthodes d'accès au spectre les enchères, la théorie des jeux, les approches de Markov et les Systèmes multi agents. Nous donnerons plus de dés taille concernant les systèmes multi agents car c'est cette technique que nous allons utiliser pour la partie application de notre travail, le dernier chapitre présenter notre contribution dans le cadre de ce RAPPORT. Nous terminons ce rapport par une conclusion générale avis que les perspectives.

# Chapitre I
# Réseaux de radio cognitive

**I.1 Introduction**

Dans ce chapitre nous nous intéresserons aux réseaux de radio cognitive avec leur fonctionnement, leur architecture et leurs différents domaines d'application.

La demande croissante de la communication sans-fil introduit le défi de l'utilisation efficace du spectre. Pour relever ce défi IEEE 802.22 est un nouveau groupe de travail du comité de normalisation WRAN. La radio cognitive est apparue comme une technologie clé, qui permet un accès opportuniste au spectre et répondre directement au besoin lié à la gestion de l'environnement du terminal radio.

**I.2 Radio logicielle (software radio)**

Le concept de radio logicielle est issu des travaux de Joseph Mitola en 1991 pour définir une classe de radiocommunication configurable utilisant des techniques de traitement numérique du signal sur des circuits numérique programmables.

La radio logicielle permettra de configurer les systèmes radio par logiciel et de manière dynamique, la bande passante du signal, la modulation et l'accès au réseau, la fréquence porteuse sont réalisés sous forme matériel. Les radios logicielles modernes mettent également en œuvre des fonctions cryptographiques, codage correcteur d'erreur, de la vidéo ou des données.

Les interfaces radio peuvent augmentées la vitesse de communication, pour différents usages, voire même de façon simultanée, et d'interopérabilité entre systèmes et de répondre au besoin croissant de performance des équipements embarqués.

Nous distinguons plusieurs niveaux de progrès dans ce domaine: la radio logicielle (software defined radio) parfois appelée radio reconfigurable ou radio intelligente est aujourd'hui un des sujets chauds parmi les multiples activités radioamateur. Les contraintes reprogrammables d'une même plateforme matérielle pour différentes couche physique, réduction des temps de développement et des qualifications requises. La figure I.1 montre l'évolution de la radio matérielle à la radio logicielle.

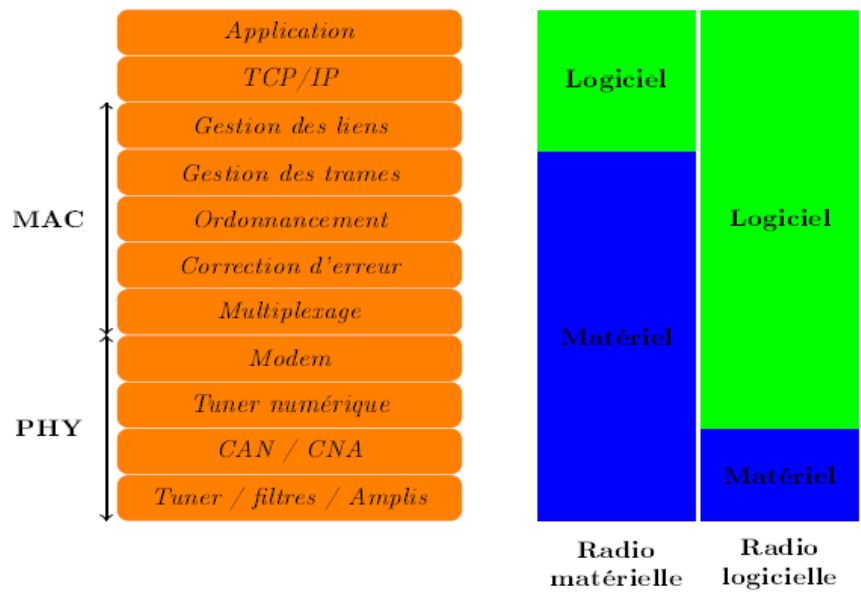

**Figure I.1 :** Evolution de la radio matérielle à la radio logicielle [1].

**I.2.1 Radio logicielle restreinte (SDR)**

La radio logicielle restreinte est un système de radiocommunication qui définie une technologie qui apporte la flexibilité et permet de résoudre des problèmes de la gestion dynamique du spectre. L'utilisation de cette technologie permet de nouvelle fonctionnalité sans fil hétérogènes (SDR idéal) avec les capacités qui peuvent être ajoutées aux systèmes radio existants sans nécessiter de matériel nouveau.

## I.3 Radio cognitive
### I.3.1 Historique
Le concept de la radio cognitive a été proposé par Joseph Mitola III lors d'un séminaire à KTH (Royal Institute of Technology), en 1998 et publié dans un certain nombre de publications (par exemple, plus tard dans un article de Mitola et Gerald Q. Maguire, Jr en 1999 [2], [3]) et dans la thèse de Doctorat de Mitola en 2000 [4]. Ce travail a été destiné à décrire les radios intelligentes qui peuvent prendre des décisions de manière autonome en utilisant les informations recueillies sur l'environnement RF grâce à un modèle basé sur le raisonnement, et peut également apprendre et de planifier en fonction de leur expérience passée. De toute évidence, un tel niveau d'intelligence exige que la radio doit être conscient de soi, ainsi que le contenu et sensibles au contexte [5]. La figure I.2 représente les étapes du développements dans la radio cognitive.

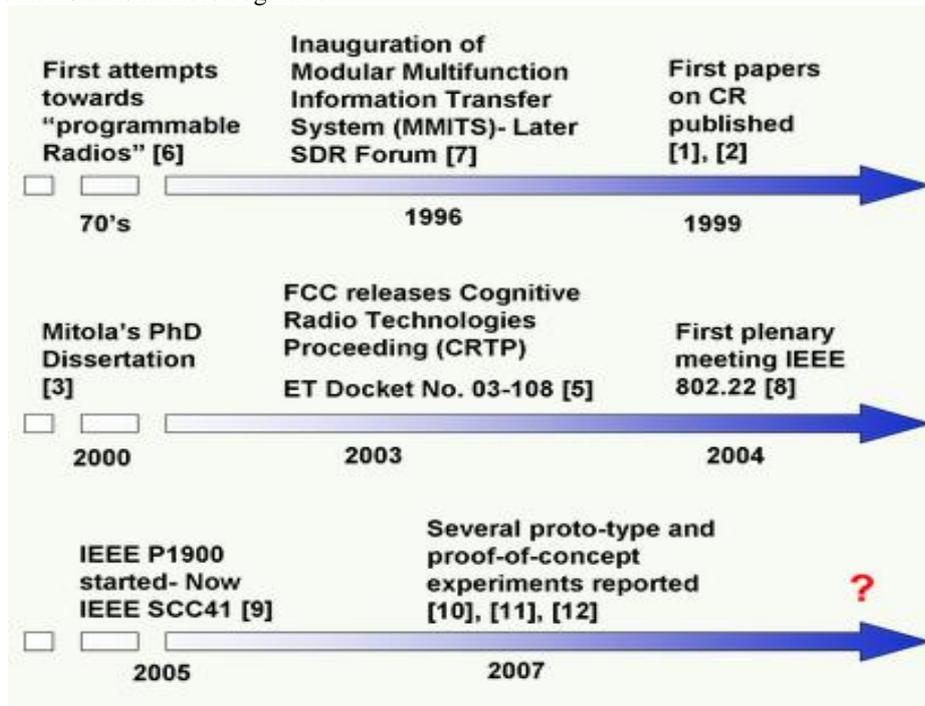

**Figure I.2 :** Développements dans la radio cognitive [5].

### I.3.2 Définition
Une radio opportuniste est un système sans fil dans laquelle un émetteur/récepteur conçu pour utiliser les meilleurs canaux de communication dans son voisinage. Cette radio ayant la capacité de reconnaître son cadre d'utilisation et de détecter automatiquement les canaux qui sont disponibles et ceux qui ne le sont pas dans le spectre. L'utilisation des fréquences radio du spectre permet de minimiser les interférences entre les terminaux.

L'objectif de la radio cognitive est donc d'ouvrir les bandes licenciées à ces terminaux secondaires sans perturber les communications des utilisateurs primaires qui sont seuls censés l'occuper [6].

### I.3.3 Relation entre radio cognitive et radio logicielle restreinte
La radio cognitive (RC) traite un ensemble des caractéristique comme la capacité d'adaptation (bande passante, puissante, modulation, fréquence porteuse) dépends de l'environnement radio, les besoin de l'utilisateur, l'état du réseau, ...etc.

La radio logicielle (RL) associée une flexibilité logicielle capable d'offrir le changement de contexte (changement de service, de traitement) et introduit les notions de reconfigurabilité dit dynamique, implique une reconfiguration sans interruption de service, et l'aspect d'adaptabilité ou encore de modularité.

La relation entre la RC et la SDR est définie par un modèle simple qui est représenté dans la figure I.3. Dans ce modèle, représenter les éléments de la RC entourent le support SDR et le cognitive engine qui est nécessaire pour la prise de décision et l'apprentissage de l'environnement radio dans un nuage qui exploite efficacement les ressources disponibles.

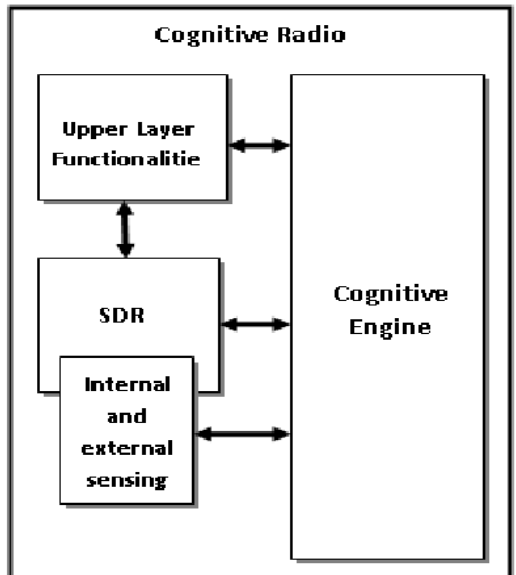

**Figure I.3 :** Relation entre radio cognitive et radio logicielle restreinte [7].

**I.3.4  Architecture**
Introduite par J. Mitola, l'architecture d'une radio cognitive est un ensemble spécifique de six composantes réalise une série de fonctions et de service, elle est illustrée dans la Figure I.4.
- La perception sensorielle (Sensory Perception : SP) de l'utilisateur qui inclut l'interface haptique (du toucher), acoustique, la vidéo et les fonctions de détection et de la perception.
- Les capteurs de l'environnement local (emplacement, température, accéléromètre, etc.).
- Les applications système (les services médias indépendants comme un jeu en réseau).
- Les fonctions SDR (qui incluent la détection RF et les applications radio de la SDR).
- Les fonctions de la cognition (pour les systèmes de contrôle, de planification, d'apprentissage).
- Les fonctions locales effectrices (synthèse de la parole, du texte, des graphiques et des affiches multimédias) [7] [8].

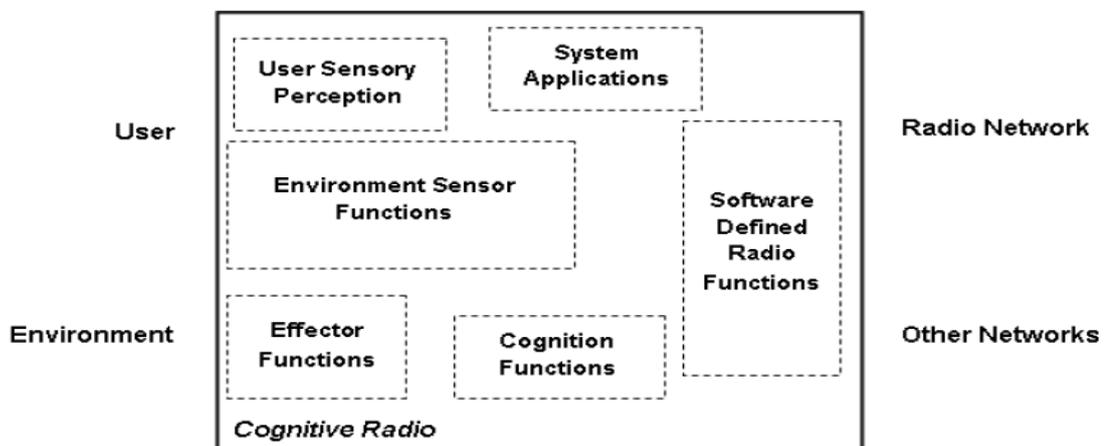

**Figure I.4 :** Architecture de la radio cognitive.

L'architecture de la RC utilise les protocoles d'adaptation de la couche MAC pour établir des interfaces entre l'émetteur/récepteur SDR et les applications sans fil (l'état du réseau), la couche physique dont le RF est définie par logiciel et traite le signal par des algorithmes intelligents. La figure I.5 présente les protocoles utilisés par la radio cognitive.

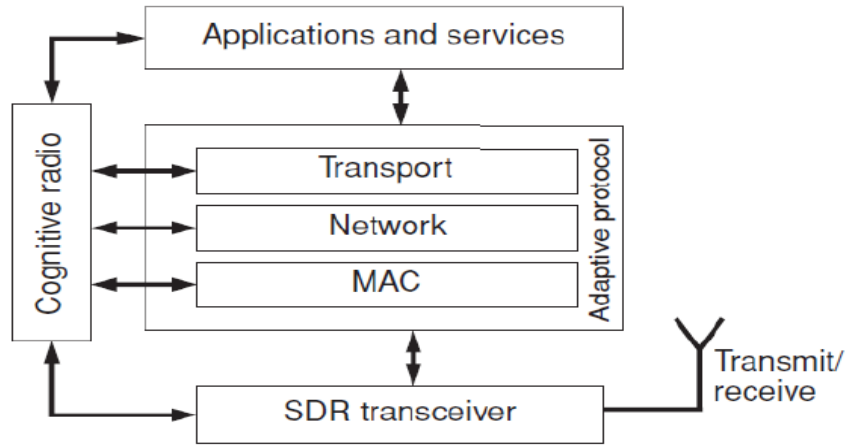

**Figure I.5 :** Protocoles utilisés par la radio cognitive [7].

### I.3.5  Cycle de cognition
Un cycle de cognition peut contrôler la circulation de l'information dans l'environnement radio est illustré dans la Figure I.6 [4].

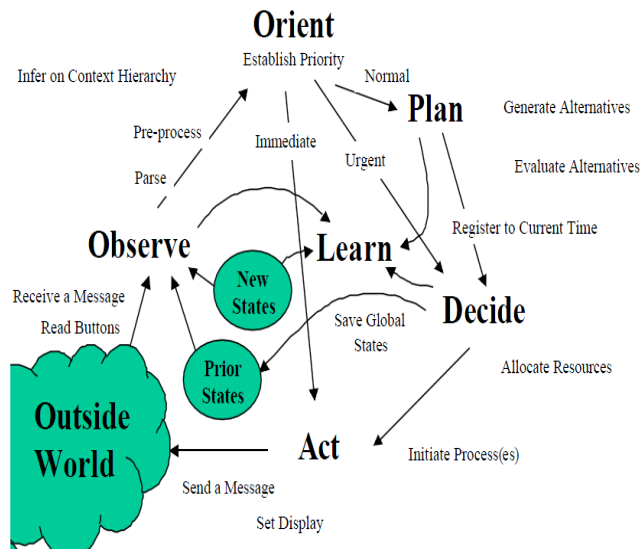
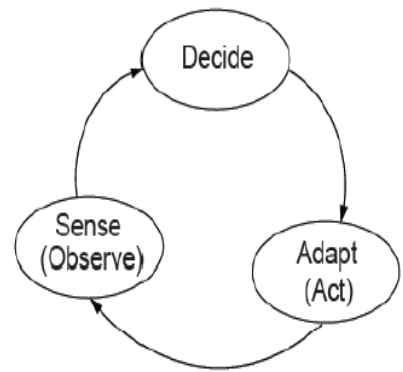

**Figure I.6 :** cycle de cognition.                **Figure I.7 :** cycle de cognition simplifié.

La radio cognitive s'oriente vers la détermination de la priorité associée à des stimuli. Une coupure d'alimentation puissante par exemple peut invoquer directement un acte. Toutefois, un message entrant du réseau serait normalement traité par la génération d'un plan (chemin "normal"). Les modèles formels de la causalité [9] sont intégrés dans les outils de planification, cette phase devrait également inclure le raisonnement. La radio pourrait avoir le choix pour alerter l'utilisateur à un message entrant ou de reporter jusqu'à plus tard l'interruption. L'apprentissage est une fonction d'observations et de décisions. Par exemple, les états internes préalables et actuelles peuvent être comparés avec les attentes à apprendre sur l'efficacité d'un mode de communication [4].

### I.3.6  Composantes de la radio cognitive
Les différentes composantes d'un émetteur/récepteur radio cognitive qui mettent en œuvre ces fonctionnalités sont présentées dans la figure I.8.

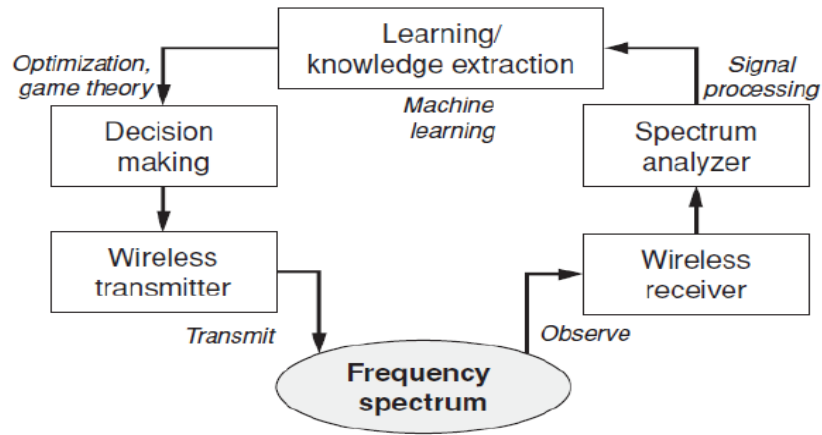

**Figure I.8 :** Composantes de la radio cognitive [10].

- 
- **Emetteur/récepteur :**
  Un récepteur RC utilisé aussi pour observer l'activité sur le spectre de fréquence (détection de spectre). Les paramètres de ce composant peuvent être modifiés dynamiquement comme dicté par les protocoles de couche supérieure.
- **Analyse de spectre (Spectrum analyser) :**
  L'analyseur de spectre diverses techniques de traitement du signal utilisées pour obtenir des informations sur l'utilisation du spectre, détecte la signature d'un utilisateur primaire et trouver les espaces blancs du spectre pour les utilisateurs secondaire et s'assurer que la transmission de donnée d'un PU n'est pas perturbée si un SU décide d'accéder au spectre.
- **Apprentissage et extraction de connaissance (Learning/ knowledge extraction) :**
  L'apprentissage et l'extraction de connaissance utilisent les algorithmes d'apprentissage et les informations sur l'utilisation du spectre pour comprendre l'environnement ambiant RF (le comportement des utilisateurs sous licence comme PU).
- **Prise de décision (Decision making) :**
  La décision sur l'accès au spectre doit être faite après que la connaissance de l'utilisation du spectre soit disponible. La décision optimale dépend du comportement coopératif ou compétitif des utilisateurs secondaires SU et dépend du milieu environnement ambiant RF. Les méthodes d'optimisation stochastique (le processus de décision de Markov) utilisée pour modéliser et résoudre le problème d'accès au spectre dans un environnement RC.

**I.3.7 Fonctions de la radio cognitive**

❖ **Détection du spectre (Spectrum sensing)**

C'est la fonctionnalité de base, détection des portions du spectre vides par détection de signaux d'utilisateurs sous licence, elle consiste à :
- Détecter le spectre non utilisé ;
- Partager le spectre sans interférence avec d'autre utilisateur.

L'objectif de cette fonction est de détecter des interférences pour obtenir l'état du spectre (libre/occupé) par SU.

❖ **Gestion du spectre (Spectrum management)**

Capturer la bande de fréquence disponible pour répondre aux besoins de communication des utilisateurs avec des fonctions classées comme suit :

- **Analyse du spectre**

Analyser les résultats de la détection du spectre pour estimer la qualité du spectre (la disponibilité des espaces blancs du spectre, durée moyenne).

- **Décision sur le spectre**

Prise de la décision pour l'accès au spectre dépend des résultats de l'analyse du spectre. Partage des portions du spectre détectés avec d'autres utilisateurs ou coexistant avec eux sur la même bande par des techniques comme d'optimisation stochastique. Dans un système RC coopératifs/non coopératifs, il existe deux utilisateurs (PU et SU) qui peuvent être influé sur l'accès au spectre.

Dans un environnement non-coopératif, chaque utilisateur a son propre objectif, tandis que dans un environnement coopératif, tous les utilisateurs peuvent collaborer pour atteindre un seul objectif. Par exemple, plusieurs utilisateurs secondaires peuvent entrer en compétition les uns avec les autres pour accéder au spectre (par exemple, O1, O2, O3, O4 dans la figure I.9) de sorte que leur débit individuel soit maximisé. Au cours de cette opération, tous veillent à ce que l'interférence causée à l'utilisateur primaire est maintenue en dessous de la limite de température de brouillage correspondante. La théorie des jeux est l'un des outils les plus appropriés pour obtenir la solution d'équilibre pour ce problème dans un tel scénario.

Dans un environnement coopératif, les RCs coopèrent les unes avec les autres pour prendre une décision pour accéder au spectre et maximiser une fonction objective commune en tenant compte des contraintes. Dans un tel scénario, un contrôleur central peut coordonner la gestion du spectre [7].

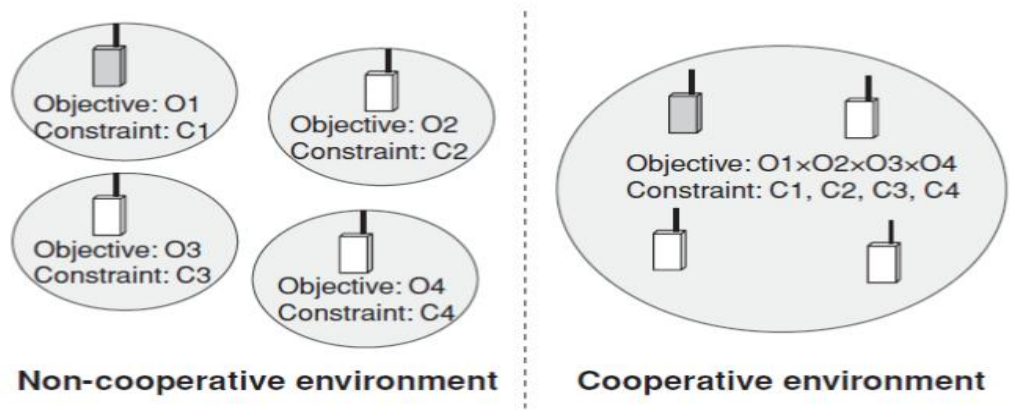

**Figure I.9 :** Accès au spectre coopératif et non-coopératif.

❖ **Mobilité du spectre (Spectrum mobility)**

La mobilité des terminaux de la radio cognitive permettant à changer sa bande de fréquence, donc l'utilisation du spectre de manière dynamique.
- Recherche des meilleures bandes de fréquence disponible ;
- Auto-coexistence et synchronisation ;
  Laissant la partie du spectre lorsque l'utilisateur sous licence d'héritage veut l'utiliser de façon isolée, sinon un utilisateur primaire peut partager la fréquence avec un utilisateur secondaire avec une certaine contrainte comme la durée.

## I.4 Domaines d'application de la radio cognitive

Parmi les domaines d'application de la radio cognitive, on peut citer :

❖ **Les réseaux sans fil de prochaine génération :**

La radio opportuniste (RC) est apparue comme une technologie clé pour la prochaine génération des réseaux sans fil hétérogènes.

❖ **Coexistence de différentes technologies sans fil :**

La RC est une solution qui fournit la coexistence de différentes technologies sans fil, pour relever ce défi, l'IEEE 802.22 est un nouveau groupe de travail du comité de normalisation WRAN dont l'objectif est d'utiliser efficacement les bandes de fréquence TV.

❖ **Services de cyber santé (eHealth services) :**

Dans ce cas, les équipements et appareils utilisent la transmission RF. L'utilisation du spectre doit être choisi avec soin pour éviter toute interférence, donc les concepts de la radio peuvent être appliqué.

❖ **Réseaux d'urgence :**

Les réseaux d'urgence peuvent profiter des concepts de la radio pour fournir la fiabilité et la flexibilité de communication sans fil.

❖ **Réseaux militaire :**

Les paramètres de la radiocommunication sans fil peuvent être adapté de manière dynamique dépend du temps et de l'emplacement.

## I.5 Conclusion

Dans ce chapitre, nous avons présenté les caractéristiques principales de la radio cognitive, nous avant introduit plusieurs concepts comme la radio logicielle, la SDR, l'architecture de RC, le cycle de cognition, les composantes de la RC, les fonctions de la RC et les domaines d'application.

# Chapitre II
# Accès dynamique au spectre dans le cadre de la radio cognitive

## II.1 Introduction

La demande croissante des communications sans fil et l'utilisation du spectre statique représente un problème majeur dans l'environnement radio. Pour résoudre le problème, les radios cognitives utilisent l'accès dynamique au spectre. Quatre techniques sont utilisées : les enchères, la théorie des jeux, les approches de Markov et les Systèmes multi agents. Nous donnerons plus de détaille concernant les systèmes multi agents car c'est cette technique que nous allons utiliser pour la partie application de notre travail.

## II.2 Techniques d'accès dynamique au spectre

### II.2.1 Les enchères

La théorie des enchères est basée sur des règles simples, qui permettent de faciliter la répartition dans les bandes de fréquence à tous les utilisateurs (PU et SU).

Il existe plusieurs formes d'enchères, notamment :
- ❖ Enchères anglaises : enchère publique au premier prix ascendante.
- ❖ Enchères hollandaises : enchère publique au premier prix descendante.
- ❖ Enchères scellée au premier prix.
- ❖ Enchères scellée au second prix : (Vickrey).

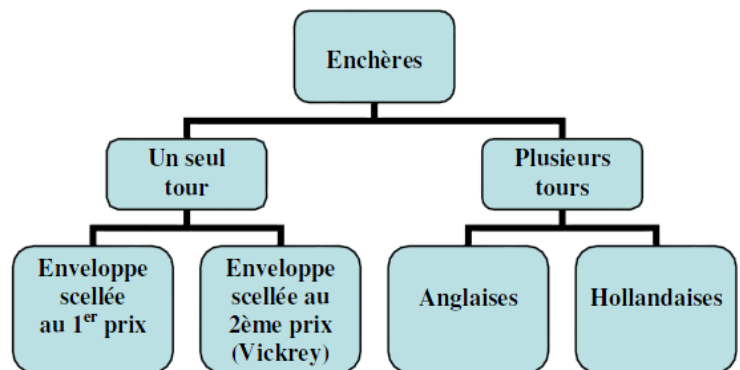

**Figure II.1 :** Organigramme représentant les types d'enchères [11].

### II.2.2 La théorie des jeux

Les jeux sont généralement divisés en deux types : jeux compétitifs et jeux coopératifs.

- **Jeux compétitifs :** tous les joueurs sont préoccupés par tous les gains globaux et ils ne sont pas très inquiets de leur gain personnel. Certains travaux récents [12] [13] utilisent la théorie des jeux coopératifs pour réduire la puissance de transmission des utilisateurs secondaires afin d'éviter de générer des interférences avec les transmissions des utilisateurs primaires.
- **Jeux coopératifs :** chaque utilisateur est principalement préoccupé par son gain personnel et donc toutes ses décisions sont prises de manière compétitive et égoïste. Dans la littérature existante, nous avons constaté que les concepts théoriques du jeu ont été largement utilisés pour des attributions de fréquences dans les réseaux RC [14] [15] [16], où lorsque les utilisateurs primaires et secondaires participent à un jeu, ils ont un comportement rationnel pour choisir les stratégies qui maximisent leurs propres gains [17].

### II.2.3 Les approche de Markov

Les approches de la théorie des jeux ne modélisent pas l'interaction entre les utilisateurs secondaires et primaires pour l'accès au spectre. Cette modélisation peut être réalisée en utilisant efficacement les chaines de Markov [18]. Une chaine de Markov est une suite de variables aléatoires qui permet de modéliser l'évolution dynamique d'un système aléatoire. La propriété fondamentale des chaines de Markov est que son évolution future ne dépend du passé qu'au travers de sa valeur actuelle. Autrement, dans le cas de la RC, cette méthode ne se contente pas du résultat seulement comme les autres méthodes mais permet également de modéliser l'interaction entre les utilisateurs (PU et SU) [19].

### II.2.4 Les Systèmes Multi Agents

L'approche classique de l'intelligence artificielle (IA), modélise le comportement intelligent d'un seul agent. Le passage du comportement individuel au comportement social pour combler les limites de l'I.A. classique à résoudre des problèmes complexes. Pour cela, nécessité de distribuer l'intelligence sur plusieurs agents.

L'intelligence artificielle distribuée (IAD) branche de l'I.A. classique s'intéresse à des comportements intelligent qui sont le produit de l'activité coopérative d'un ensemble entités.

L'I.A.D a introduit le concept de système multi-agents qui portent sur le modèle produits par les interactions d'agents dont les caractéristiques sont : La coopération, la coordination et la communication.

Les SMA sont extensibles et adaptatifs, ce qui permet d'ajouter ou d'enlever un agent du système sans causer de problèmes [33] [34]. Dans les réseaux sans fil, nous pouvons modéliser les nœuds RC comme étant des agents où à chaque fois qu'ils changent de zones (handover) le SMA change.
Les SMA sont connus aussi par leur rapidité car les agents peuvent travailler en parallèle pour résoudre leurs problèmes [19].

### II.3  Agents
- **Qu'est ce qu'un agent ?**

Le terme d'agent est utilisé dans plusieurs domaines, il est défini selon son type d'application, une des définitions qui est considérée comme l'une des premières est celle de Ferber en 1995 [20].

Un agent est une entité réelle ou virtuelle, capable de le percevoir et d'agir sur elle-même et sur l'environnement, de communication avec d'autres agents, qui possède un comportement autonome, lequel peut être vu comme la conséquence de ses connaissances, de ses interactions avec d'autres agents. Il peut appartenir à plusieurs organisations.

- **Les caractéristiques d'un agent**

Un agent est caractérisé par :
- La nature : un agent est une entité physique ou virtuelle.
- L'autonomie : un agent est indépendant de l'utilisateur et des autres agents.
- l'environnement : c'est l'espace dans lequel va agir l'agent, il peut se réduire au réseau constitué par l'ensemble des agents.
- La capacité de représentation : l'agent peut avoir une vision très locale de son environnement mais il peut aussi avoir une représentation plus large de cet environnement et notamment des agents qui l'entourent.
- La communication : l'agent aura plus ou moins de capacité à communiquer avec les autres agents.
- Le raisonnement : l'agent peut être lié à un système expert ou à d'autres mécanismes de raisonnements plus ou moins complexes.
- L'apprentissage : un agent aura plus ou moins tendance à retirer, stocker et réutiliser des informations extraites ou reçus de son environnement.
- La contribution : l'agent participe plus ou moins à la résolution de problèmes ou à l'activité globale du système.
- L'efficacité : l'agent doit avoir la rapidité d'exécution et d'intervention [7].

- **Types d'agents**

On distingue deux types d'agents :
  - ➢ Agents cognitifs :

Chaque agent dispose d'une connaissance, il comprend toute l'information pour la gestion des interactions avec les autres agents et son environnement. Exemple typique : système Multi-Experts.
  - ➢ Agents réactifs :

Chaque agent possède un mécanisme de réaction aux événements qui ne prenant en compte ni des mécanismes de planification, ni une explicitation des buts. Exemple typique : systèmes multi-agents de simulation.

Le tableau suivant représente les différents points entre l'agent cognitif et l'agent réactif.

| Agent cognitif | Agent réactif |
|---|---|
| Représentation explicite de l'environnement et des autres agents. | Pas de représentation explicite de l'environnement, des connaissances. |
| Peut tenir compte de son passé. | Pas de mémoire de son histoire. |
| Agent complexe. | Comportement de type stimulus/réponse. |
| Petit nombre d'agent. | Grand nombre d'agents. |

**Tableau II.1 :** Les différents points entre l'agent cognitif et l'agent réactif.

## II.4 Systèmes Multi Agents

Un Système Multi-Agents (SMA) est un système distribué composé d'un ensemble d'entités réactifs ou cognitifs (suivant le problème traité), qui interagissent les uns avec les autres, situé dans un environnement commun. Un SMA est caractérisé ainsi :

- Chaque agent a des compétences élémentaires ou des capacités de résolution des problèmes limités (ainsi, chaque agent a un point de vue partiel).
- Il n'y a aucun contrôle global du système multi-agents.
- Les données sont décentralisées (car si un agent tombe en panne, le système continue de fonctionner).
- Ajout de nouvelles composantes (agents) à un SMA ne pose aucun problème ce qui explique leur extensibilité et modularité.
- Apprentissage / adaptation.
- Interactions : interaction au niveau supérieur pour résolution distribuée des problèmes, organisation (centralisée vers décentralisée), communication (langage et protocole de communication) et coordination (agents individualistes, coopératifs et compétitifs).

La figure II.2 représente la synthèse de SMA dans la littérature selon quatre dimensions (agents, environnement, interaction et organisation).

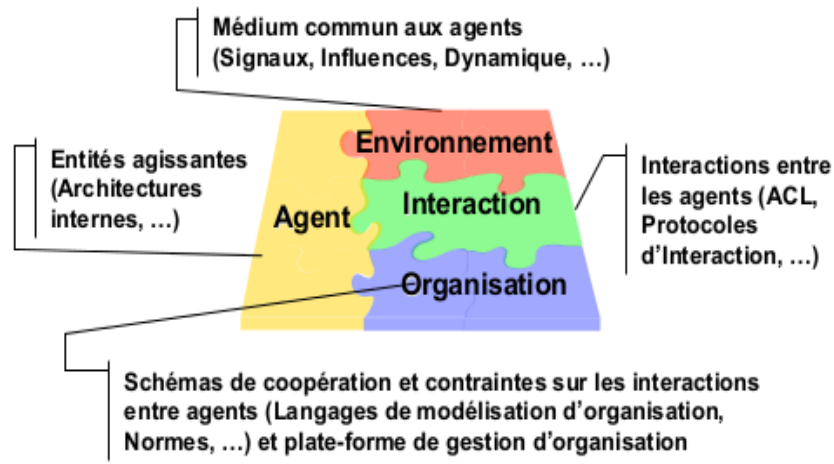

**Figure II.2 :** Synthèse du SMA.

## II.5 SMA et Radio Cognitive

L'association des SMA avec la RC assure un futur remarquable pour la gestion optimale des fréquences (en comparaison avec les techniques de contrôle rigides proposées par les opérateurs de télécommunications). Dans le cas de l'utilisation des bandes sans licence, le terminal RC doit coordonner et coopérer pour un usage meilleur du spectre sans causer d'interférences.

Dans [7], les auteurs proposent une architecture basée sur les agents où chaque terminal RC est équipé d'un agent intelligent, il y a des modules pour collecter les informations à propos de l'environnement radio et bien sur les informations collectées seront stockées dans une base de connaissance partagée qui sera consultée par tous les agents. L'approche proposée est basée sur les SMA coopératifs (les agents ont des intérêts en commun). Ils collaborent en partageant leurs connaissances pour augmenter leur gain individuel ainsi que collectif.

Des agents sont déployés sur les terminaux RC des PU et des SU et coopèrent entre eux dans les travaux proposés dans [21] [22]. Par SMA coopératif, on veut dire que les agents PU échangent des t-uples de messages dans le but de s'améliorer eux-mêmes ainsi que le voisinage des agents SU. Ils proposent que les SU doivent prendre leur décision en se basant sur la quantité du spectre disponible, le temps et le prix proposé par les agents PU. Et ils doivent commencer le partage du spectre dès qu'ils trouvent une offre appropriée (Sans attendre la réponse de tous les PU). En d'autres termes, l'agent SU doit envoyer des messages à l'agent PU voisin approprié, et bien sur le PU concerné doit répondre à ces agents pour faire un accord sur le partage du spectre. Et bien sur après la fin de l'utilisation du spectre, le SU doit payer le PU.

Pour rendre les systèmes de RC pratiques, il faut que plusieurs réseaux RC coexistent entre eux. Cependant, ceci peut générer des interférences. Les auteurs de [23] pensent que pour remédier à ce problème, les SU peuvent coopérer pour détecter le spectre aussi bien que pour le partager sans causer d'interférences pour le PU. Pour cela, ils proposent des schémas pour protéger les PU des interférences en contrôlant la puissance de transmission du terminal cognitif.

Dans [24] [25], les auteurs proposent une coopération entre les PUs et les SUs et entre les SUs seulement. Des agents sont déployés sur les terminaux des utilisateurs pour coopérer et aboutir à des contrats régissant l'allocation du spectre. Les agents SU coexistent et coopèrent avec les agents PU dans un environnement RC Ad hoc en utilisant des messages et des mécanismes de prise de décision. Vu que les comportements internes des agents sont coopératifs et désintéressés, ça leur permet de maximiser la fonction d'utilité des autres agents sans ajouter de coût conséquent en termes de messages échangés.

Cependant, l'allocation des ressources est un enjeu important dans les systèmes de RC. Il peut être fait en effectuant la négociation parmi les utilisateurs secondaires. Dans [26] les auteurs proposent un modèle basé sur les agents pour la négociation du spectre dans un réseau RC. Mais au lieu de négocier le spectre directement entre des PUs et des SUs, un agent courtier est inclus. Ce qui veut dire que l'équipement du PU ou du SU ne nécessite pas une grande intelligence vu qu'il n'a pas besoin d'effectuer la détection du spectre ou autre chose. L'objectif de cette négociation est de maximiser les bénéfices et les profits des agents pour satisfaire le SU. Les auteurs ont proposé deux situations, la première utilise un seul agent qui va exploiter et dominer le réseau, et dans la deuxième, il va y avoir plusieurs agents en concurrence.

Le SMA contient plusieurs agents intelligents en interaction entre eux. Chaque agent peut faire la détection et l'apprentissage. L'agent peut sélectionner les comportements basés sur l'information locale et tenter de maximiser les performances globales du système. Dans [27], les auteurs ont décrit une nouvelle approche basée sur l'apprentissage par renforcement multi-agent qui est utilisée sur des réseaux RC ad-hoc avec contrôle décentralisé. En d'autres termes, ils ont mis en place plusieurs scénarios de RC et ils affectent à chaque cas une récompense ou une pénalité. Les résultats de cette approche ont montré qu'avec cette méthode, le réseau peut converger à un partage équitable du spectre et bien sur elle permet de réduire les interférences avec les utilisateurs primaires PU [28].

La figure III.3 représente une architecture de gestion des ressources pour plusieurs WLAN en utilisant des systèmes multi-agents. Les agents sont situés à l'intérieur chaque point d'accès (AP) et d'interagir avec d'autres agents dans son voisins. La zone de l'agent consiste à celles avec lesquels il a des interactions fréquentes. Ces interactions comprennent le partage des données et à la négociation sur affectations de ressources. Les agents individuels agissent comme ressource radio coordinateurs et de coopérer avec les agents dans leur région de prendre soin de la gestion des ressources à travers de multiples réseaux locaux sans fil.

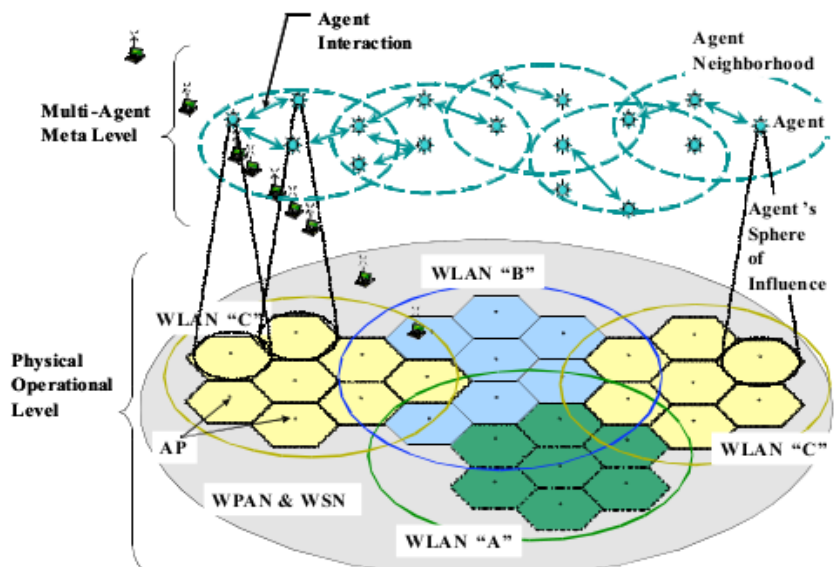

**Figure II.3 :** L'architecture de la gestion des ressources WLAN en utilisant des systèmes multi-agents [29].

**II.5.1 Protocoles de négociation**
Un protocole de négociation est l'ensemble de règles qui dirigent l'interaction. Ceci inclut les types de participants permis, les états de la négociation, les événements qui font passer d'un état à un autre et les actions valides et acceptables de la part des participants. Dans la littérature, il existe plusieurs techniques d'accès au spectre comme on a vu dans le premier chapitre et plusieurs protocoles de négociation; le tableau II.2 résume les protocoles les plus importants :

| Protocole de négociation | Description |
|---|---|
| Contract Net | Les agents coordonnent leurs activités grâce à l'établissement de contrats pour atteindre des buts spécifiques. |
| Théorie des enchères | Le terme « enchère » désigne toute technique de vente établissant une concurrence, qui a pour objectif de déterminer le futur possesseur de l'article en jeu, par des offres successives. |
| Négociation heuristique | Les agents doivent fournir des réactions plus utiles aux propositions qu'ils reçoivent, ces réactions peuvent prendre la forme d'une critique ou d'une contre-proposition (proposition refusée ou modifiée). |
| Négociation par argumentation | Un agent peut essayer de persuader un autre agent de répondre favorablement à sa proposition en cherchant des arguments qui identifient de nouvelles occasions, créent de nouvelles occasions ou modifient les critères d'évaluation. |

**Tableau II.2 :** Protocoles de négociation [11].

**II.5 Conclusion**

Un système multi-agents s'adapte mieux à la réalité des environnements complexes que l'intelligence artificielle classique.

Le principe du SMA est de faire travailler ensemble des entités pour proposer des solutions réactives et robustes à des problèmes complexes, ce qui réduit le temps de résolution vu la vitesse utilisée qui est due principalement au parallélisme.
Dans le chapitre suivant, nous proposons une solution basées sur la négociation en SMA dans le cadre de la radio cognitive pour une gestion efficace du spectre.

# Chapitre III
# Négociation de spectre : simulation et évaluation

**III.1  Introduction**

Dans un environnement RC, il y a deux types d'utilisateurs : un utilisateur dit primaire (PU) qui pourra à tout moment utiliser ses bandes de fréquence car il possède une licence appropriée. Par contre, l'utilisateur dit secondaire (SU) pourra accéder à des bandes de fréquence qu'il trouve libres, c'est-à-dire, non occupées coté PU à condition que ce dernier accepte de coopération avec lui sans causer d interférences. Dans ce type de scénario, un problème de l'encombrement causé par le manque de ressources disponibles pour les SUs est rencontré.

Dans notre scénario, nous supposons l'existante d'un seul SU et de plusieurs PUs. Le SU à un besoin application exprimé en termes de canaux et chaque PU possède en certain nombre de canaux libres, qu'il accepte de partager avec le SU.

Notre objectif est de déployer un agent par SU et un agent par chaque PU et d'implémenter une solution basée sur la négociation dans la cadre de ce SMA. La négociation doit répond à un but précis du SU. Nous avons utilisé la plate forme JADE (Java Agent Development Framework) pour l'implémenter de notre solution.

**III.2  Topologie du réseau utilisé**

Dans ce travail, nous proposons d'utiliser une architecture de réseau ad hoc, car ce type de réseau est capable de s'organiser de manière autonome sans infrastructure fixe. La figure III.1 illustre la topologie du réseau utilisé, un environnement RC ad hoc composants d'un seul SU et de plusieurs PUs.

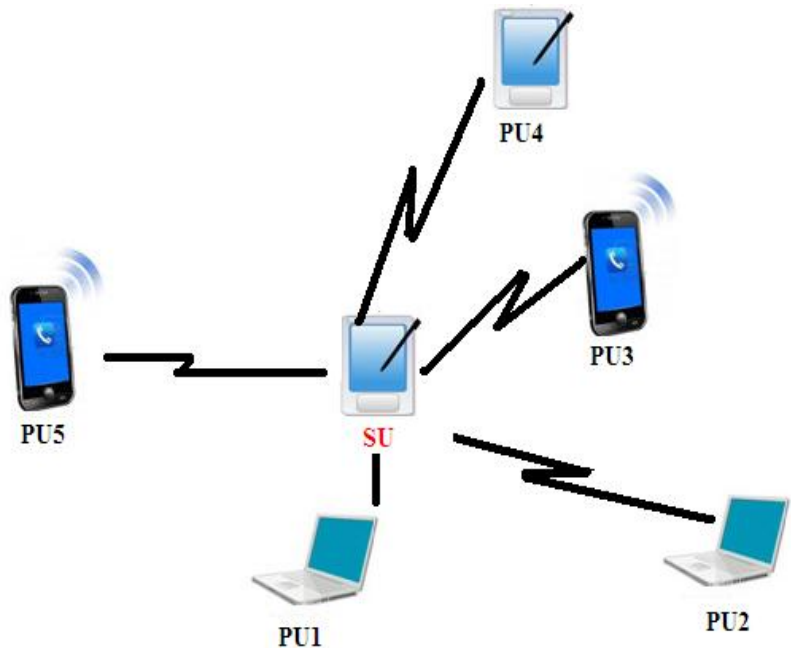

**Figure III.1 :** Topologie du réseau (mode ad hoc)**.**

**III.3  Présentation du scénario**

Pour notre application, nous nous sommes basées sur un type particulier de négociation « un à plusieurs », exactement, un SU qui va entamer une négociation avec plusieurs PUs, par exemple 5 PUs. La figure 2 donne une idée sur un éventuel scénario. Le SU a un besoin applicatif en termes de canaux, donc il va contacter chaque PU afin de connaître à la fois les canaux disponibles chez lui mais aussi le prix qu'il propose pour ses canaux. En fonction de ce qu'il reçoit de la part de tous les PUs, le SU sera en mesure de choisir la meilleure offre disponible pour lui.

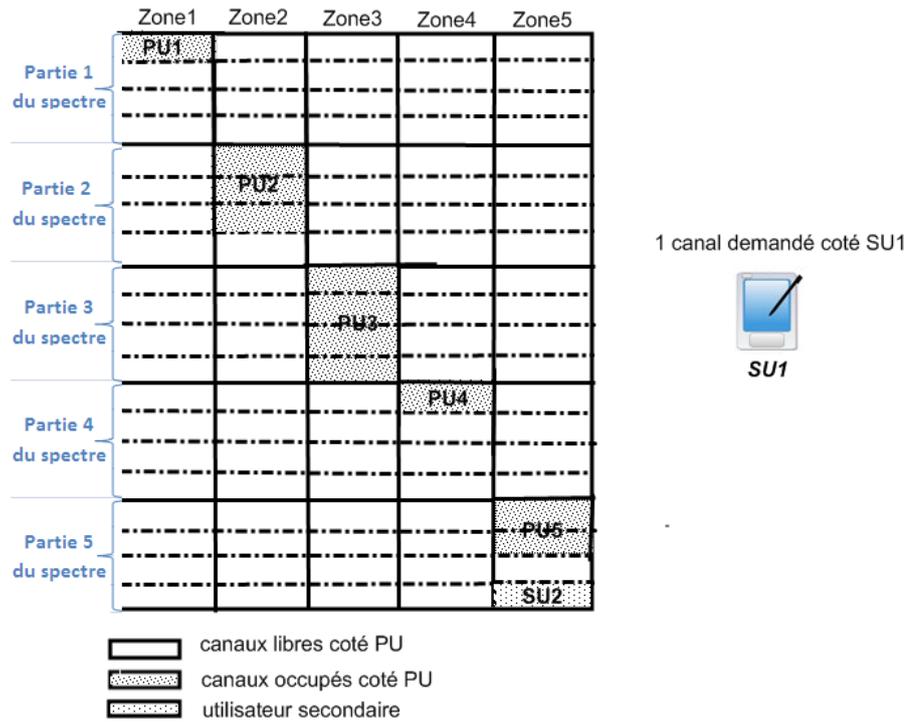

**Figure III.2 :** Scénario proposé.

### III.4  Outils utilisés
#### III.4.1  JADE

JADE (Java Agent Development Framework) est une plateforme multi agents en java developpée par Gruppo Telecom Italia en 1997, outil open source qui répond à la norme FIPA [30]. Une application jade est une plateforme déployée sur une ou plusieurs machines, la plateforme héberge un ensemble des entités de type « cognitifs », identifiés de manière unique, pouvant communiquer avec les autre agents de façon bidirectionnelle (langage de communication FIPA ACL). Chaque agent s'exécute dans un conteneur (container) qui lui fournit son environnement d'exécution.

- Architecture globale de la plateforme jade :
- AMS : (Agent Management System) agent qui supervise les autres agents et l'accès à la plateforme, pages blanches.
- ACC : (Agent Communication Chanel) agent qui fournit la route pour les interactions entre agents dans et en dehors de la plateforme.
- DF : (Director Facilitator) agent qui fournit un service de pages jaunes.
- Interface graphique :
- Jade GUI

Permet de contrôler les agents disponibles.

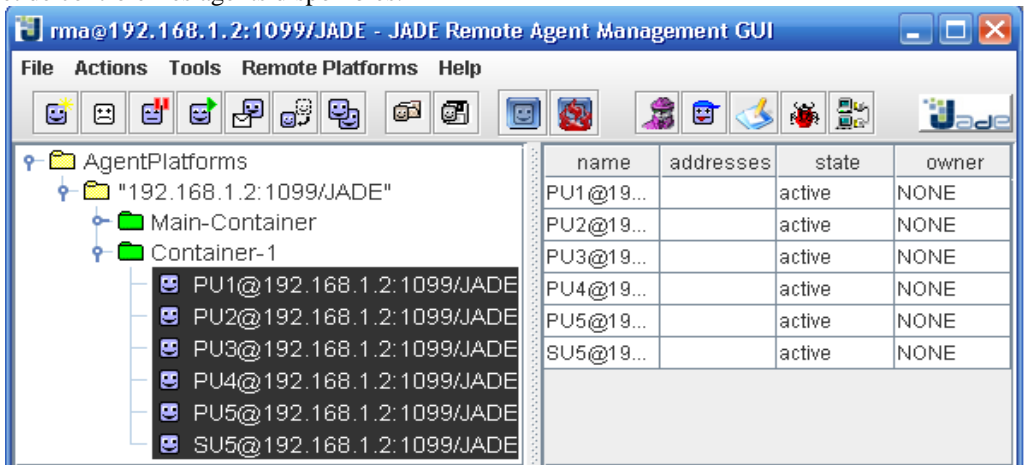

**Figure III.3 :** Interface graphique (jade GUI).

### III.4.2 Jfreechart

Jfreechart est une API java permettant de créer des graphiques et de diagramme de très bonne qualité. Il est conçu pour être inclus dans des applications, applets, servlets et JSP. Cette API est open source et sous licence LGPL (Lesser General Public Licence), ce qui permet à JfreeChart d'être utilisé dans des applications libres ou propriétaires.

### III.5 Code JADE des agents (PU et SU)

Dans ce qui suit, nous noterons :
Nbc : le nombre de canaux demandés par le SU.
NbcLibre : le nombre de canaux libres coté PU.
prix : le prix unitaire d'un canal coté PU.
tab : tableau de taille 6, tab[i] représente le prix proposé par PUi.
Le code suivant représente une partie du code JADE lié au SU.

**Code du SU**

```
//*** Début de la négociation : SU envoie aux PUs le nombre de canaux nécessaires pour son application ***//
Nbc=2
protected void setup() {
for (int i=1 ; i<= 5 ; i++){
    ACLMessage msg1 = new ACLMessage(ACLMessage.REQUEST);
    msg1.addReceiver(new AID("PU"+i, AID.ISLOCALNAME));
    msg1.setContent(""+Nbc);
        send(msg1)

//********* Traitement des réponses reçus de tous les PUs coté SU *******//
addBehaviour(new CyclicBehaviour(this)  {
                public void action() {
                        ACLMessage msge = blockingReceive();
                            {  if ((msge != null)){
                                String[] msg = msge.getContent().split(" ");
                                    tab[1] = new Integer([0]);
                                   remplir() ;} }
    else  block();   }}
}) ;
}
 protected void remplir() {
        int min = tabrecu[1];
                for (int i=2; i<= 5 ; i++){
  if (tabrecu[i]<= min){
min=tabrecu[i];
                    }              }
            if (min != 10000){
//******** Agent PU"+i+" propose le meilleur prix : " +min +" DA **********//
  int y=20000;

//********** si la demande satisfaite coté SU, le SU envoie au PU un message confirm ********//
            ACLMessage msg = new ACLMessage(ACLMessage.CONFIRM);
                    msg.setContent(""+y);
                    msg.addReceiver( new AID("PU"+i, AID.ISLOCALNAME));
                    send(msg);   }

//******* sinon si la demande n'est pas satisfaite coté SU, le SU n'envoie rien au PU *******//
                        }
```

Le code suivant représente une partie du code JADE lié au PU.

**Code du PU**

```
//********* Traitement de la demande du SU coté PU *******//
Nbc=3
   protected void setup() {
addBehaviour(new CyclicBehaviour(this) {
                        public void action() {
     block();
```

```
ACLMessage msg=receive();
ACLMessage mssage = new ACLMessage(ACLMessage.INFORM);
ACLMessage max = new ACLMessage(ACLMessage.REFUSE);
ACLMessage message = new ACLMessage(ACLMessage.ACCEPT_PROPOSAL); {
   if (msg != null){
  String t=msg.getContent(); // reçu le msg
  int i=Integer.parseInt(t);

//*** "+NbcLibre+" canaux libres sont disponible coté PUi: "+prix+" DA est le prix unitaire envoyé a
l'agent SU ***//
prix=300
if (i <= NbcLibre) {
   mssage.addReceiver(new AID("SU", AID.ISLOCALNAME));
   mssage.setContent(""+prix);
   send(mssage);
      }

//*** "+NbcLibre+" canaux libres sont disponible coté PUi, pas assez de canaux pour satisfaire la
demande du SU ***//
insuffi=1000
 else if (i >  NbcLibre) {
  max.addReceiver(new AID("SU", AID.ISLOCALNAME));
  max.setContent(""+insuffi);
  send(max);
      }

//***** le PU envoie au SU un message de type accept_proposal ******//
   else {     message.addReceiver(new AID("SU", AID.ISLOCALNAME));
           message.setContent("");
           send(message);
}}
}) ;
}
```

## III.6  Résultat obtenu

Nous avons développé une interface basée sur swing de javax et awt de java, nous utilisons deux champs de saisie de type JTextField, le premier pour entrer le nombre de canaux demandés par le SU et le deuxième pour entrer le nombre de PUs (Nb). Trois bouton de type JButton sont aussi utilisé, le bouton « Lancer les agents » pour démarrer l'ensemble des agents PUs avec cardinalité Nb, « Lancer sniffer » pour voir les interactions entre les agents et le troisième bouton « Valider » pour  tracer le graphe designer les graphes, nous utilisons aussi une zone pour afficher la trace de l'exécution à l'aide d'un JTextArea .

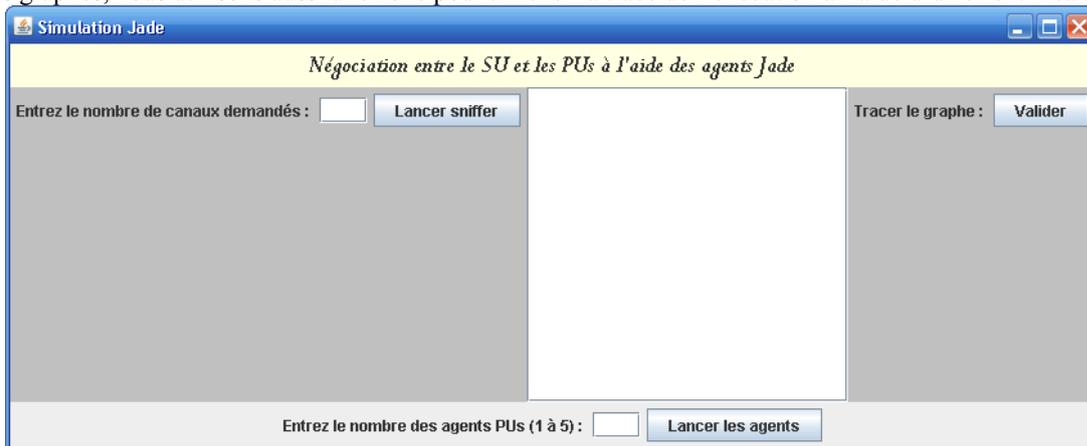

**Figure III.4 :** Interface pour lancer la simulation jade.

## III.6.1  Sniffer Agent

Nous avons implémenté trois scénario différentes en gardant le même jeu de données mais en chargement le nombre de canaux demandés par le SU à chaque fois. Dans la première simulation le SU demande 1 canal, dans la deuxième le SU demandés 3 canaux et dans la troisième le SU demandés 5 canaux. Le jeu de données que nous avons utilise est comme suit.
PU1 (1, 270) : PU1 à 1 canal libre et il propose un prix unitaire de 270.

PU2 (2, 230) : PU2 à 2 canaux libres et il propose un prix unitaire de 230.
PU3 (3, 320) : PU3 à 3 canaux libres et il propose un prix unitaire de 320.
PU4 (4, 250) : PU4 à 4 canaux libres et il propose un prix unitaire de 250.
PU5 (3, 340) : PU5 à 3 canaux libres et il propose un prix unitaire de 340.
Pour la première simulation, tous les Pus peuvent satisfais le SU mais l'offre la plus intéressent est coté PU2. L'interface de simulation dans ce cas est le suivant :

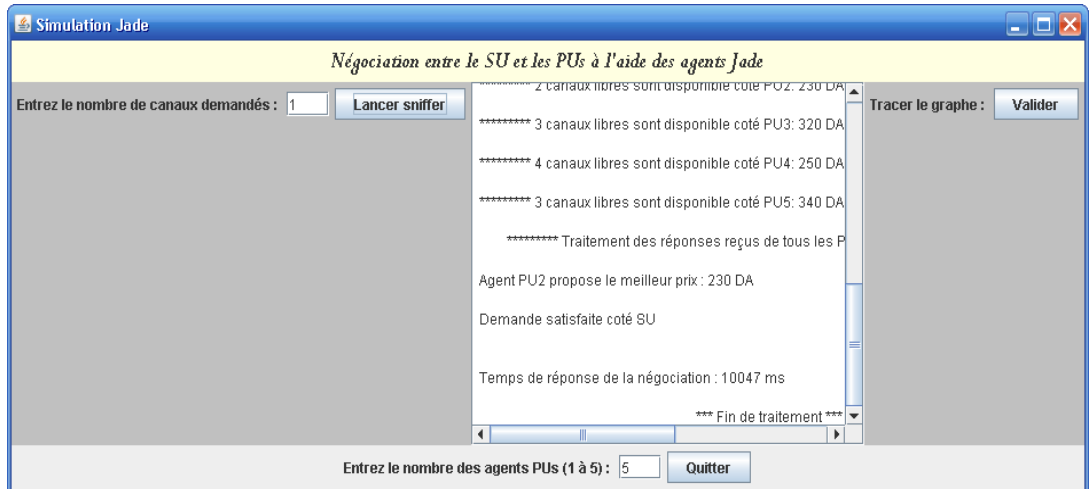

**Figure III.5 :** Interface pour entrer l'ensemble de données.

Le sniffer dans ce cas est comme suit :

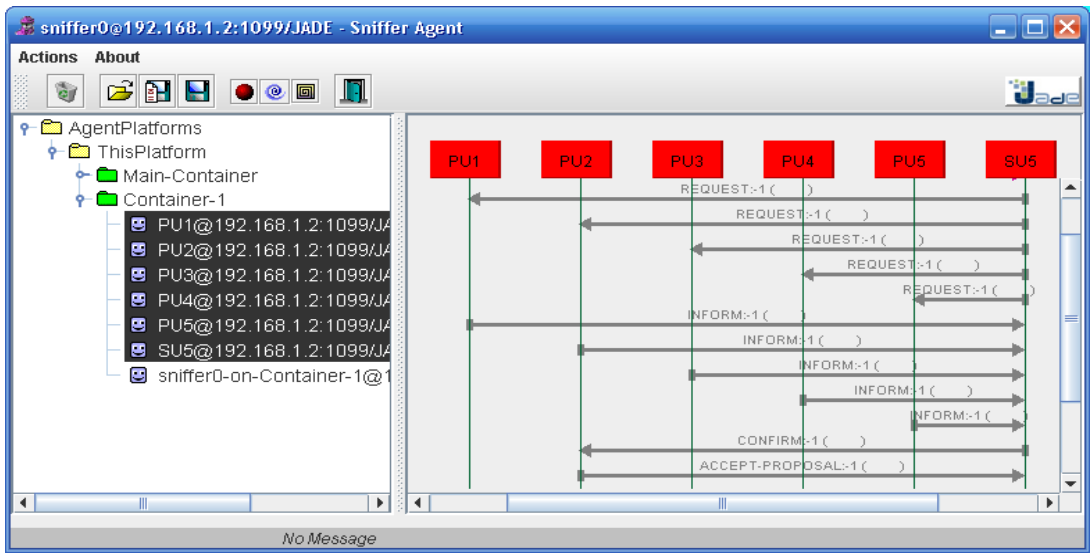

**Figure III.6 :** Sniffer avec 1 canal demandé coté SU.

Pour la deuxième simulation, PU1 et PU2 ne peuvent pas satisfais le SU, l'offre la plus intéressent est coté PU4.

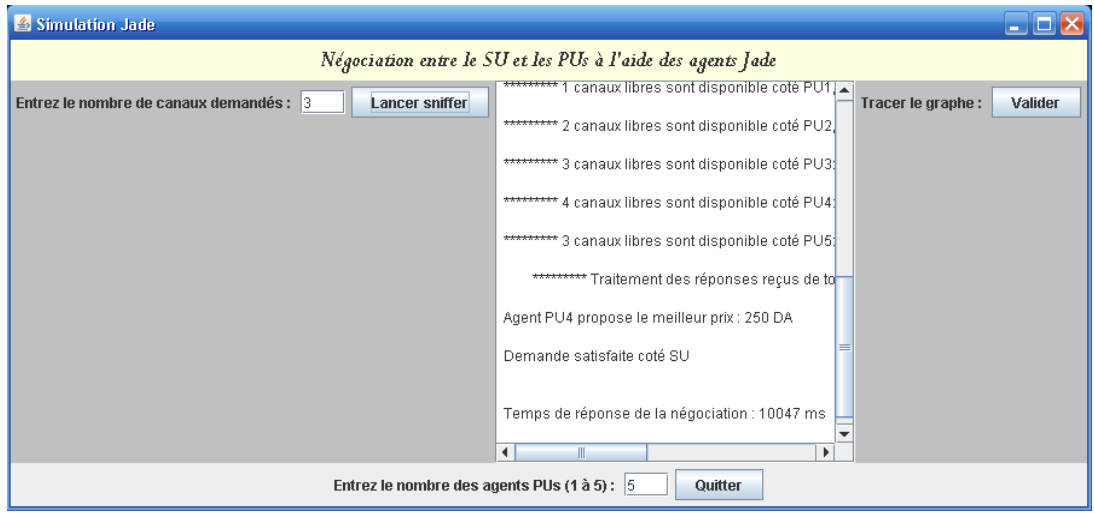

**Figure III.7 :** Interface pour entrer l'ensemble de données.

Le sniffer est comme suit :

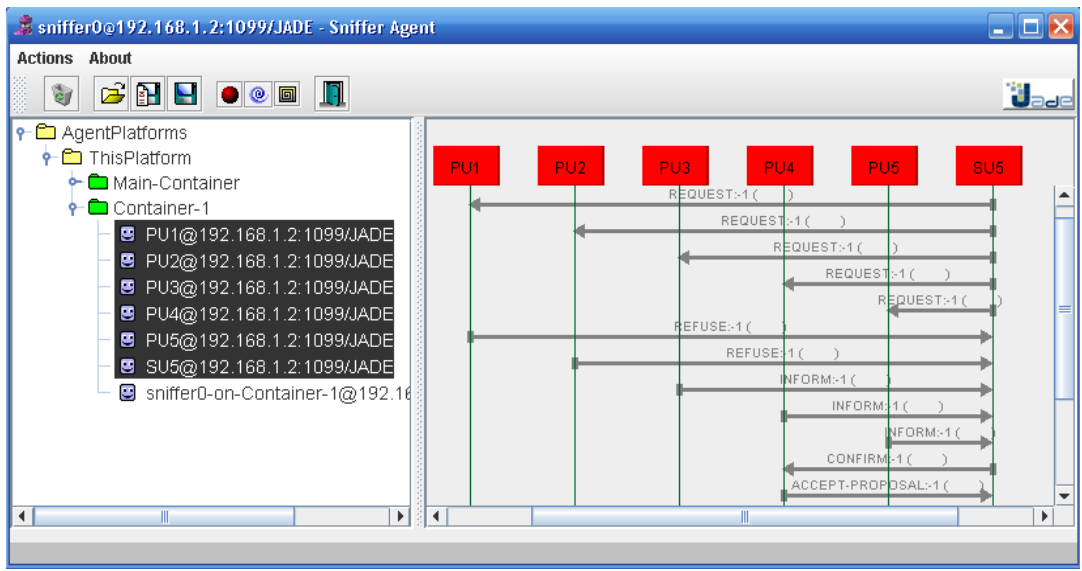

**Figure III.8 :** Sniffer avec 3 canaux demandés coté SU.

Pour la troisième simulation, il n'y a aucun PU qui peut satisfais le SU.

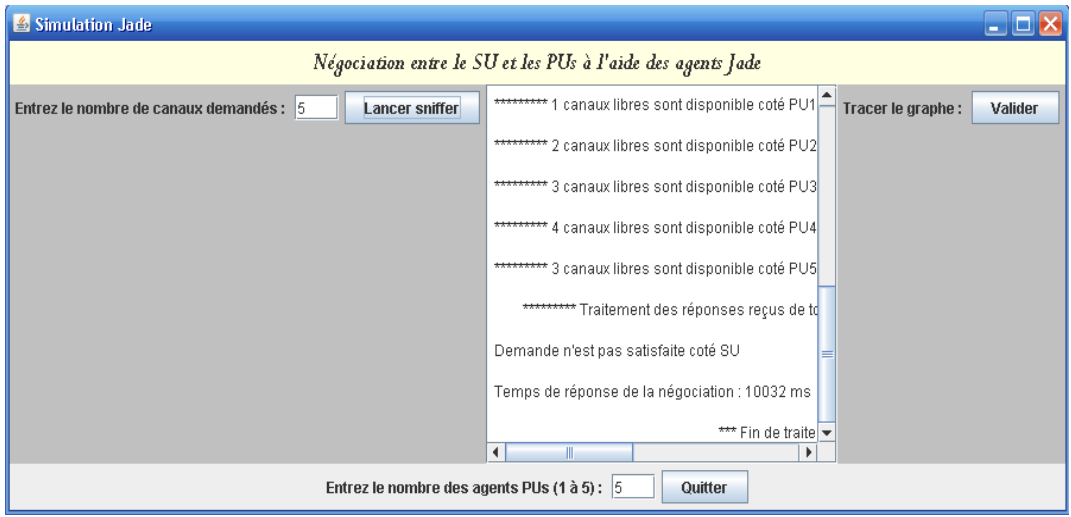

**Figure III.9 :** Interface pour entrer l'ensemble de données.

Le sniffer est comme suit :

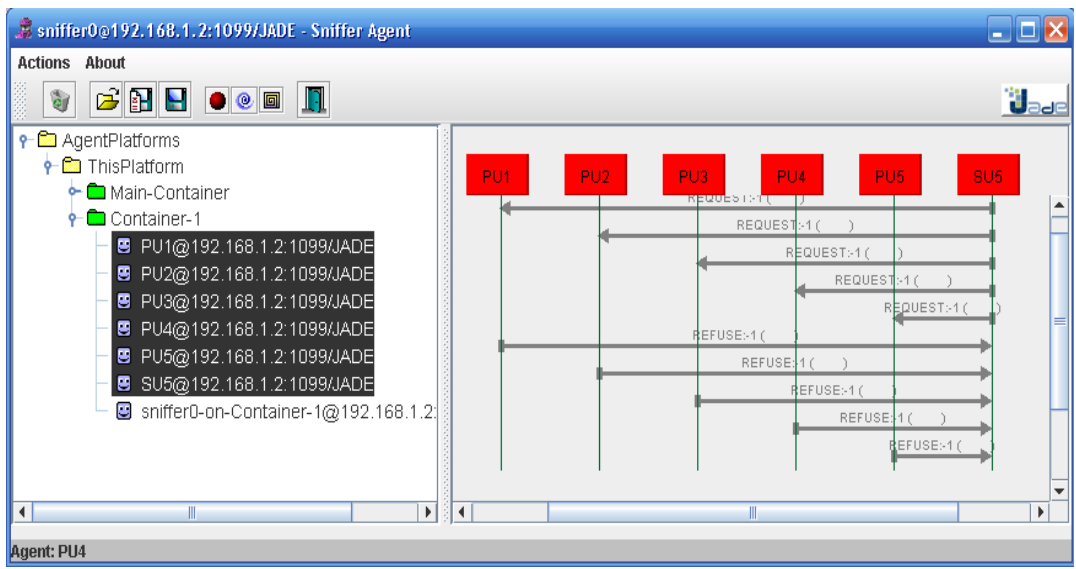

**Figure III.10 :** Sniffer avec 5 canaux demandés coté SU.

Dans ce qui suivant, nous souhaiteras mesurer l'impact du nombre de PUs sur le temps de réponse coté SU, pour se faire, on fixe le nombre de canaux coté SU à 3 (Nbc=3) et on utilise le même jeu de données précédente :
PU1 (1, 270), PU2 (2, 230), PU3 (3, 320), PU4 (4, 250) et PU5 (3, 340).

Le temps de réponse mesuré coté SU est indiqué dans le tableau III.1.

| Nombre de PUs | Temps de réponse coté SU |
|---|---|
| 1 | 3 |
| 2 | 6 |
| 3 | 7 |
| 4 | 9 |
| 5 | 12 |

**Tableau III.1 :** Temps de réponse coté SU / Nombre de PUs.

Le graphe suivant montre le résultat obtenu.

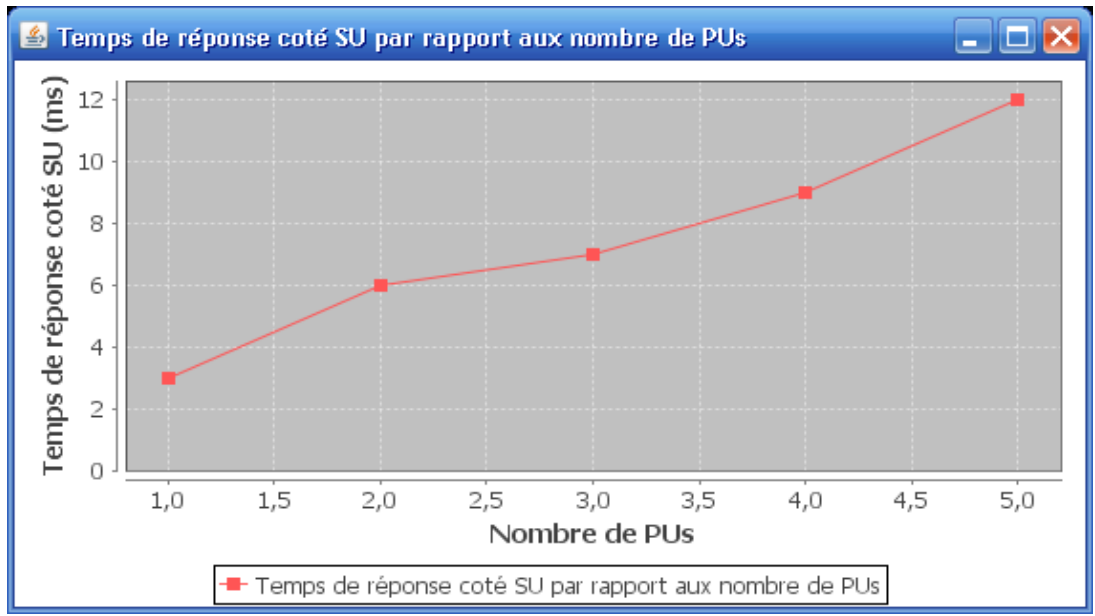

**Figure III.11 :** L'impact du nombre de PUs sur le temps de réponse coté SU (ms).

La figure 11 montre que le temps de négociation augmente pour le SU en négociant avec plus de PU. C'est logique car il y aura plus de réponse à traiter. Avec 5 PUs, le temps de négociation est de 12 ms par contre avec 2 PUs il est uniquement à 6 ms.
Notre objectif maintenant est d'évaluer l'importance de la négociation par rapport au prix payé par le SU. Pour cela, nous supposons que le SU a besoin de 2 canaux (Nbc=2).
Le tableau III.2 montre le jeu de données pour représente le cas d'échec et le cas de réussite de la négociation.

| L'échec de la négociation | | La réussite de la négociation | |
|---|---|---|---|
| **NbcLibre** | **prix** | **NbcLibre** | **prix** |
| 2 | 500 | 1 | 500 |
| 1 | 400 | 1 | 120 |
| 1 | 240 | 1 | 300 |
| 1 | 220 | 1 | 320 |
| 1 | 120 | 2 | 100 |

**Tableau III.2 :** Le jeu de données pour l'échec et la réussite de la négociation.

Un échec de négociation représente le cas où le SU va négocier pour rien, ici, il paye 500, la valeur du premier PU. Pas négociation avec les autre PUs n'a rien apporté. Un sucée de la négociation et quand celle-ci est rentable, il va négocier aves tous les PUs et l'offre les plus intéressante n'est pas celle du premier PU.
Le tableau III.3 représente les résultats obtenus.

| Taux de réussite de la négociation (%) | Prix payé par le SU |
|---|---|
| 0 | 10*500 = 5000 |
| 10 | 100*1+9*500 = 4600 |
| 20 | 100*2+8*500 = 4200 |
| 30 | 100*3+7*500 = 3800 |
| 40 | 100*4+6*500 = 3400 |
| 50 | 100*5+5*500 = 3000 |
| 60 | 100*6+4*500 = 2600 |
| 70 | 100*7+3*500 = 2200 |
| 80 | 100*8+2*500 = 1800 |
| 90 | 100*9+1*500 = 1400 |
| 100 | 100*10 = 1000 |

**Tableau III.3 :** Le jeu de données.

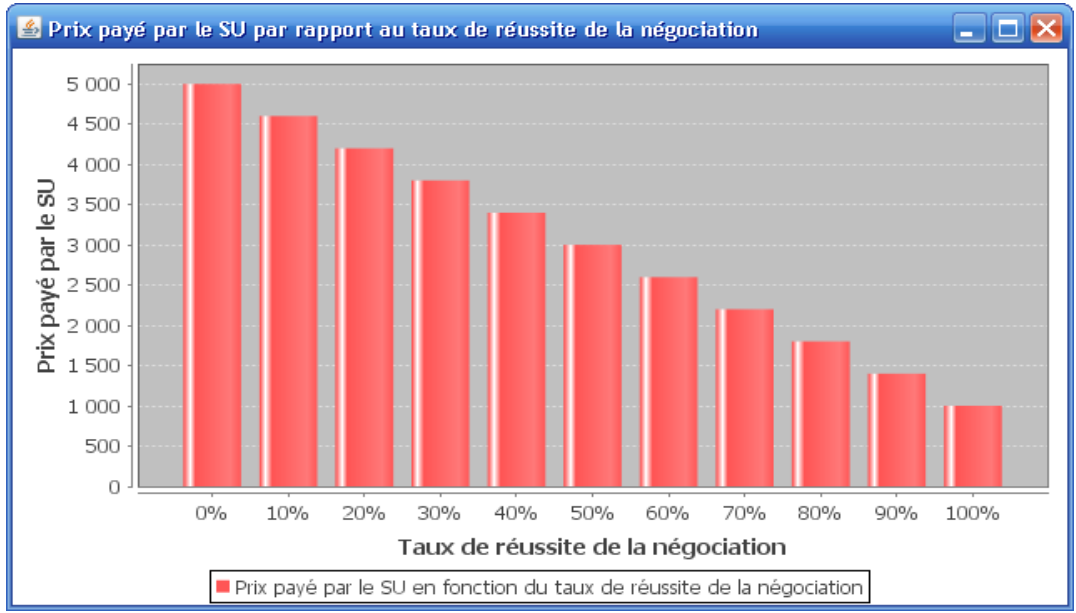

**Figure III.12 :** L'impact du taux de réussite de la négociation sur le prix payé par le SU.

L'histogramme précédent montre l'intérêt de la négociation pour le SU, la négociation prend plus de temps pour la mettre en œuvre surtout dans le cas de plusieurs PU, mais elle est toujours intéressante pour le SU car elle lui permet de trouver l'offre la plus intéressante, donc il a beaucoup de chance pour réduire ses dépenses par rapport au cas où il ne fait pas de négociation.

### III.7  Conclusion

Dans ce chapitre, nous avons présenté notre solution basée sur la négociation entre agents pour le partage de spectre dans le cadre d'un réseau de type radio cognitive. La solution proposée a été implémentée à l'aide de l'outil JADE, les résultats obtenus montrent que la négociation est intéressante pour le SU car elle lui permet de trouver la meilleure offre disponible malgré le temps de négociation qui est en légers augmentation en fonction du nombre de PU.

**Conclusion générale**

Une radio cognitive est un système sans fil ayant la capacité de détecter automatiquement les canaux qui sont disponibles et ceux qui ne le sont pas dans le spectre. L'utilisation des fréquences radio du spectre permet de minimiser les interférences entre les utilisateurs.

Les interactions entre les agents fonction critique dans les systèmes de radiocommunication, nous avons ainsi exploré certains aspects du problème de l'encombrement du spectre pour ensuite le résoudre en introduisant les systèmes multi agents.

Un SMA s'adapte mieux à la réalité des environnements complexes que l'intelligence artificielle classique. La négociation est une solution pour que les terminaux secondaires puissent négocier des ressources libres en termes de canaux cotée utilisateurs primaires.

Notre solution proposée dans le cadre de ce contexte est validée a travers la simulation avec JADE, nous avons montré que avec la négociation un SU peut trouver les offres qui s'adaptent avec ses besoins en réduisant ce qu'il doit payer aux PUs. Comme perspective à ce travail, on peut l'étendre dans le cadre de mobilité où le SU va rencontrer un nombre assez important de PU, Dans ce cas un SU doit négocier pendant sa mobilité avec tous les PUs rencontrés dans sa zone de mobilité [31] [32].